\documentclass[aps,preprintnumbers,amsmath,amssymb,prb,superscriptaddress,longbibliography,twocolumn]{revtex4-2}
\usepackage{dcolumn}% Align table columns on decimal point
\usepackage{color}
\usepackage[caption=false]{subfig}
\usepackage{feynmp}
\usepackage{feynmp-auto}
\usepackage{systeme,mathtools}
\usepackage{stmaryrd}
\usepackage{easyReview}

\def\comment#1{}

\newcommand{\nc}{\newcommand}
\nc{\beq}{\begin{eqnarray}}
	\nc{\eeq}{\end{eqnarray}}
\nc{\scs}{\scriptstyle}
\nc{\setval}{\fmfset{wiggly_len}{3mm} \fmfset{arrow_len}{1.5mm}
	\fmfset{arrow_ang}{13} \fmfset{dash_len}{1.5mm}\fmfpen{0.125mm}
	\fmfset{dot_size}{2thick}}

\usepackage{bm,latexsym,mathrsfs,enumerate,color}
\usepackage[mathcal]{euscript}
\usepackage[breaklinks=true,unicode=true,urlcolor = blue,colorlinks = true,citecolor = blue,linkcolor = blue]{hyperref}
\usepackage{graphicx}
\usepackage{todonotes}
\usepackage{wrapfig}
\usepackage{easyReview}

\renewcommand{\vec}[1]{\bm{#1}}

\def\slashchar#1{\setbox0=\hbox{$#1$}           % set a box for #1
	\dimen0=\wd0                                 % and get its size
	\setbox1=\hbox{/} \dimen1=\wd1               % get size of /
	\ifdim\dimen0>\dimen1                        % #1 is bigger
	\rlap{\hbox to \dimen0{\hfil/\hfil}}      % so center / in box
	#1                                        % and print #1
	\else                                        % / is bigger
	\rlap{\hbox to \dimen1{\hfil$#1$\hfil}}   % so center #1
	/                                         % and print /
	\fi}                                         %

\DeclareMathAlphabet\mathbfcal{OMS}{cmsy}{b}{n}

\def\nablab{{\mbox{\boldmath $\nabla$}}}
\def\varphib{{\mbox{\boldmath $\varphi$}}}

\def\Omegab{{\mbox{\boldmath $\Omega$}}}

\begin{document}
	
	\title{Chiral Meissner state in time-reversal invariant Weyl superconductors}
	
	\author{Vira Shyta}
	\affiliation{Institute for Theoretical Solid State Physics, IFW Dresden, Helmholtzstr. 20, 01069 Dresden, Germany}
	
	\author{Jeroen van den Brink}
	\affiliation{Institute for Theoretical Solid State Physics, IFW Dresden, Helmholtzstr. 20, 01069 Dresden, Germany}
	\affiliation{Institute for Theoretical Physics and W\"urzburg-Dresden Cluster of Excellence ct.qmat, TU Dresden, 01069 Dresden, Germany}
	
	\author{Flavio S. Nogueira}
	\affiliation{Institute for Theoretical Solid State Physics, IFW Dresden, Helmholtzstr. 20, 01069 Dresden, Germany}

\begin{abstract}
Weyl semimetals have nodes in their electronic structure at which electrons attain a definite chirality.
Due to the chiral anomaly, the non-conservation of charges with given chirality, the axion term appears in their effective electromagnetic action. 
We determine how this affects the properties of time-reversal invariant Weyl {\it superconductors} (SCs) in the London regime.
For type II SCs the axion coupling generates magnetic $B$-fields transverse to vortices, which become unstable
at a critical coupling so that a transition into type I SC ensues.
In this regime an applied $B$-field not only decays inside the SC within the London penetration depth, but the axion coupling generates an additional perpendicular field.
Consequently, when penetrating into the bulk the $B$-field starts to steadily rotate away from the applied field. 
At a critical coupling the screening of the magnetic field breaks down. 
The novel chiral superconducting state that emerges has a periodically divergent susceptibility that separates onsets of chiral Meissner regimes.
The chiral anomaly thus leaves very crisp experimental signatures in structurally chiral Weyl SCs with an axion response.
\end{abstract}
	
	\maketitle
	
\section{Introduction} 

Experimentally superconductivity has been reported in a number of Weyl semimetals, both at ambient \cite{YPtBi_PhysRevB.84.220504,MoTe2,PdTe2_PhysRevLett.119.016401,PdTe2_PhysRevB.96.220506,Joseph-BKT,kuibarov2023superconducting} and high pressures \cite{Weyl-high-pressure,WTe-high-pressure}. The topological nature of Weyl semimetals \cite{Burkov-Balents_PhysRevLett.107.127205,Burkov_PhysRevB.85.165110,Burkov_PhysRevB.86.115133,Grushin_PhysRevD.86.045001,Qi_PhysRevB.90.045130,Felser-review,Burkov-review} gives hope that Majorana zero modes bounded to vortices \cite{Kobayashi-Sato_PhysRevLett.115.187001,Vortex-ZMs-Weyl-Dirac_PhysRevLett.124.257001} may be detected in the future. Another recent experimental development in the field is the observation of superconductivity in the time-reversal invariant (TRI) Weyl semimetal PtBi$_2$ \cite{Joseph_PhysRevMaterials.4.124202,Joseph-BKT,kuibarov2023superconducting}, where superconductivity is observed to be confined to the surface of the material.

Since the presence of Weyl nodes modifies electromagnetic properties of a superconducting system, we will explore here the Meissner  and magnetic vortex states of  Weyl superconductors (SCs)  in the London limit. The low-energy effective theory of Weyl semimetals is governed by the axion action \cite{Burkov-Balents_PhysRevLett.107.127205,Burkov_PhysRevB.85.165110,Burkov_PhysRevB.86.115133,Grushin_PhysRevD.86.045001,Qi_PhysRevB.90.045130,Felser-review,Burkov-review}, $S_a=\frac{\alpha}{4\pi^2}\int dt\int d^3r\vartheta(t,\vec{r}) \vec{E}\cdot\vec{B},$ where $\alpha$ is the fine-structure constant and the axion field is assumed to have the explicit form $\vartheta(t,\vec{r})=\vec{b}\cdot\vec{r}-b_0t$. Here $\vec{b}$ and $b_0$ represent the separation between Weyl nodes in momentum and energy, respectively \cite{Felser-review,Burkov-review}.  
Specifically, we will be interested in the case where time-reversal invariance holds, which leads to a net $\vec{b}=0$ due to the presence of time reversed Weyl node pairs. 
In the normal state at equilibrium Weyl nodes of opposite chirality cancel out the axion action term proportional to $b_0$ \cite{Landsteiner_2016,Arouca2022,Franz_PhysRevLett.111.027201}. This is not necessarily so in Weyl SCs, in particular when
 Weyl cones of one particular chirality are gapped out due to the superconducting pairing \cite{Beenakker_PhysRevLett.118.207701,PACHOLSKI2020168103}, as in this case the remaining gapless Weyl nodes of opposite chirality induce axion electrodynamics. This scenario is relevant in cases where the normal groundstate Weyl nodes of opposite chirality have different energies. In a TRI crystal this requires the absence of any mirror symmetry and, as inversion must be broken to produce Weyl nodes in the first place, thus a chiral crystal structure \cite{Rao-CoSi,li2022chirality}. Superconductivity in such chiral crystals permits gapping out of Weyl nodes of a single chirality in equilibrium so that the axion action term proportional to $b_0$ appears in the effective electromagnetic response.
 
In such cases the London electrodynamics of TRI Weyl SCs is captured by the following Lagrangian, 
\begin{eqnarray}
	\label{Eq:L}
	\mathcal{L}&=&\frac{\epsilon}{8\pi}\vec{E}^2-\frac{1}{8\pi}\vec{B}^2+\frac{\rho_s}{2}\left[\left(\partial_t \theta+q \phi\right)^2-(\nablab \theta-q \vec{A})^2\right]\nonumber\\&-&\frac{q^2b_0}{16\pi^2}   \vec{A}\cdot\vec{B},
\end{eqnarray} 
where $\rho_s$ is the superconducting stiffness, $q=2e$ the charge,  $\theta$ the phase of the order parameter, $ \phi$ and $\vec{A}$ are electric and vector potentials, respectively. We use units such that $\hbar=c=1$.  As we are interested in the static regime, we obtain the following generalized London equation from the Lagrangian above, 
\begin{equation}
	\label{eq:curlofB}
	\nablab \times \vec{B} = 4\pi q\rho_s(\nablab\theta-q\vec{A}) - a \vec{B},
\end{equation}
where from Eq. (\ref{Eq:L}) we identify the axion coupling constant as given by $a=q^2b_0/(2\pi)$.

 Within the London theory the superconducting current is given by  $\vec{j}_{\rm SC}=q\rho_s(\nablab\theta-q\vec{A})$, while a typical physical consequence of the axion response is the chiral magnetic effect (CME) \cite{KHARZEEV2014133,Burkov_2015,Armitage2018}, which implies that the current density contains a contribution $\vec{j}_{\rm CME}=-a\vec{B}/(4\pi)$. Therefore, in a Weyl superconductor with time-reversal invariance the total current density is given by $\vec{j}=\vec{j}_{\rm SC}+\vec{j}_{\rm CME}$, which is reflected in Eq. \eqref{eq:curlofB}.  Here we uncover a number of novel electrodynamic features that follow from the interplay between the axion CME and superconductivity in chiral Weyl semimetallic materials.

As will be shown, while the magnetic field expulsion from a superconductor is ensured by its current having a term proportional to $\vec{A}$, the CME contribution, which is linear in $\vec{B}$, leads to a rotation of the screened magnetic field. The latter behavior may be understood by first considering the non-superconducting phase. In this case $\nablab\times\vec{B}=-a\vec{B}$ and we see that $\nabla^2\vec{B}+a^2\vec{B}=0$, which yields spatially rotating magnetic field profiles. When the system becomes superconducting, the screening of the magnetic field twists as a response to the rotation induced by the CME, and the  \textit{chiral Meissner state} ensues.

We find that the electromagnetic response of Weyl SCs differs drastically depending on whether the axion coupling $a$ is above or bellow a critical value $a_c$, which is related to the superconducting stiffness. Investigating the vortex properties of type II Weyl SCs, one observes that for $a<a_c$ a component of the magnetic induction  transverse to the vortex line is generated in addition to the one directed along the vortex line. 	The rotation of the magnetic induction around the vortex results in a screening current directed not only around but also along the vortex line. However, due to the competition between the axion term and superconductivity with the increase of the ratio $a/a_c$, a transition occurs at $a_c$ from the chiral vortex state to a Meissner state without vortices. 

For type I Weyl SCs we show that the chiral Meissner state present for $a<a_c$ breaks at $a=a_c$ and the system enters a new state with no magnetic field screening. Instead of decaying, the magnetic field rotates inside the sample for $a\geq a_c$. Nevertheless, in this regime the magnetic susceptibility $\chi =-1/(4\pi)$ at quantized values of $a$ implying diamagnetic behavior.

\section{Vortex in TRI Weyl SC}

To understand the electromagnetic response of a Weyl SC, we first focus on the  fate of a single magnetic vortex. The analysis here differs significantly from previous discussions of axion electrodynamics based on the Witten effect \cite{WITTEN1979283}, where the field of the vortex induces a fractional charge at the interface between a SC and a topological insulator \cite{Nogueira_PhysRevLett.117.167002}, as well as a fractional angular momentum  \cite{Nogueira_PhysRevLett.121.227001,Nogueira_PhysRevResearch.4.013074}. The vortex axion physics discussed below does not involve the electric field and is intrinsic to TRI Weyl SCs, so proximity to a topological material needs not be assumed. 
Taking the curl of Eq. (\ref{eq:curlofB}) we obtain, 
\begin{equation}
	-\nabla^2 \vec{B}+a \nablab \times \vec{B}+M^2 \vec{B}=\frac{M^2\Phi_0}{2\pi}\Omegab,
	\label{eq:nabla2B}
\end{equation}
where $M^2=4\pi q^2\rho_s$ represents the inverse square of the London penetration depth $\lambda$,  $\Phi_0=2\pi/q$ is the elementary flux quantum, and $\Omegab=\nablab\times\nablab\theta$ is the vorticity (recall that the curl of a gradient vanishes everywhere, except where topological defects like vortices exist \cite{kleinert1989gauge}). 

\begin{figure}
	\subfloat[]{\includegraphics[width=0.5\linewidth]{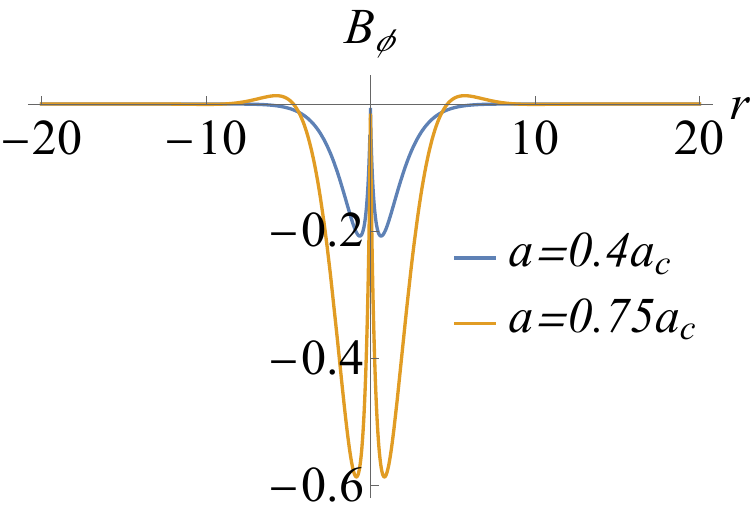}}
	\hfill
	\subfloat[]{\includegraphics[width=0.5\linewidth]{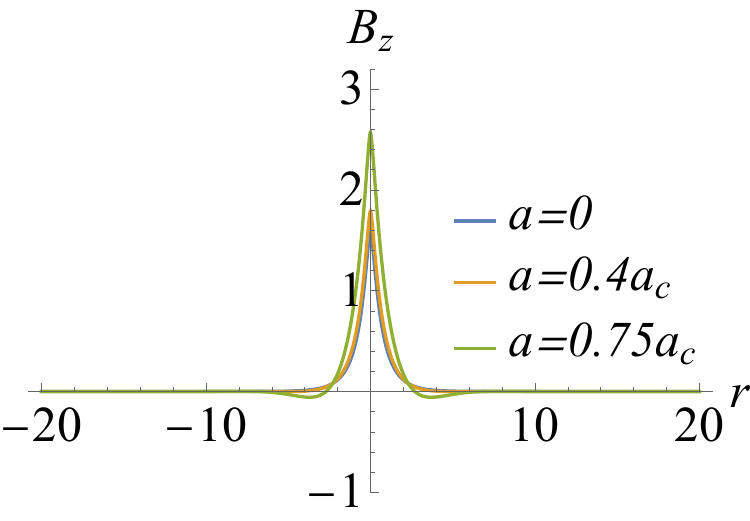}}\\
	\subfloat[$a=1.99a_c/2$]{\includegraphics[width=0.5\linewidth]{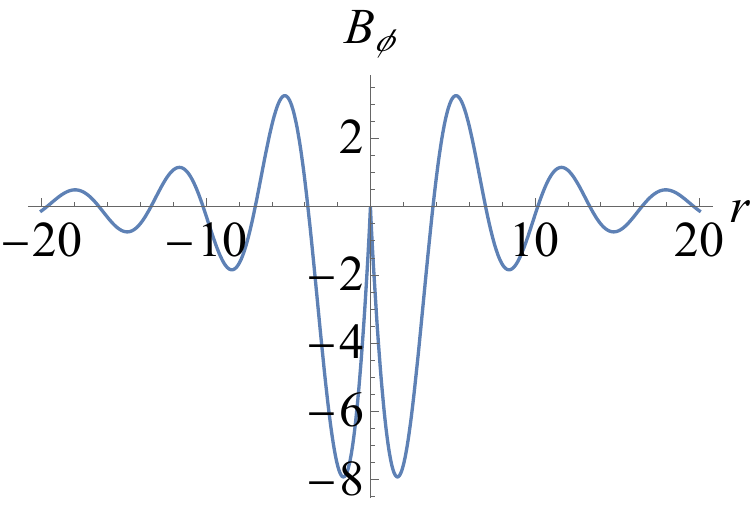}} 
	\hfill	
	\subfloat[$a=1.99a_c/2$]{\includegraphics[width=0.5\linewidth]{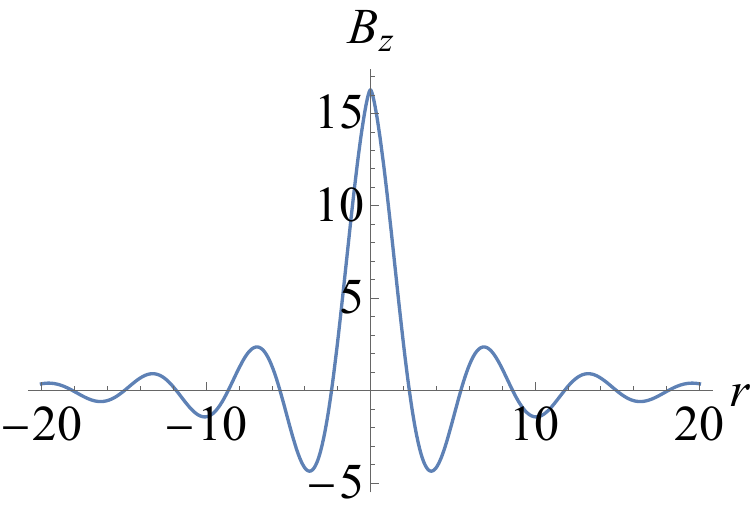}}\\
	 \subfloat[$a=0.75a_c$]{\includegraphics[width=0.5\linewidth]{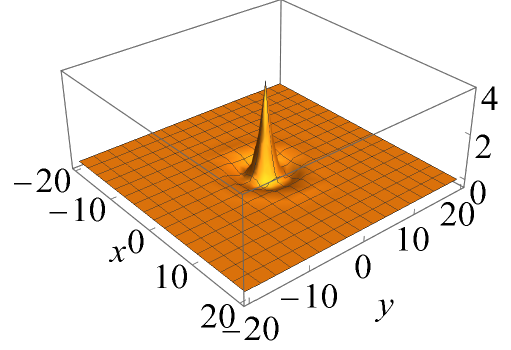}} 
	 \hfill	
	 \subfloat[$a=1.99a_c/2$]{\includegraphics[width=0.5\linewidth]{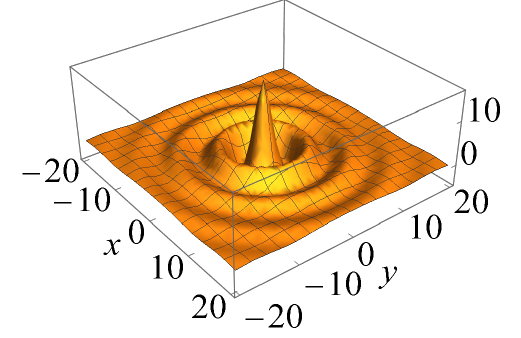}}
	\caption{Magnetic induction profiles for different values of the axion coupling $a$ [panels (a) and (b) for the field components $B_\varphi$ and $B_z$, respectively]. Fields are plotted in units of $M^2\Phi_0/(2\pi)$ against the radial coordinate $r$ in units of $M$. Panels (c) and (d) correspond to a situation where $a$ is very close to the critical value $a_c=2M$ for which the vortex solution ceases to exist. We note that as $a$ approaches $a_c$ from bellow the field profiles start to become more oscillatory. The onset of these spatial oscillations is illustrated by the three-dimensional plots for $B_z$ in panels (e) and (f).}
	\label{Fig:fig-vortex-infinite}
\end{figure}

We consider an infinite system with a single vortex line directed along the $z$-axis and with the origin located at $r=0$. This vortex configuration leads to the vorticity taking a simple form, $\vec{\Omega} = 2 \pi \delta^2(\vec{r})\hat{\vec{z}}$. The exact solution to Eq.~\eqref{eq:nabla2B} is obtained by performing a Fourier transform, which leads to
\begin{eqnarray}
	\!\!\!\!\!\! B_i(p)=\frac{2\pi M^2\Phi_0 \delta(p_z)(p^2+M^2)}{(p^2+M^2)^2-a^2p^2}
	\left(\delta_{i z}+i\frac{a\varepsilon_{i z k} p_k}{p^2+M^2} \right)\!\!, 
	\label{eq:nBi}
\end{eqnarray}
and yields in real space
%
%\begin{equation}
	$\vec{B}(\vec{r})=B_\varphi(r)\hat{\varphib}+B_z(r)\hat{\vec{z}}$,
%\end{equation}
with
\begin{equation}
	\label{Eq:Bphi}
	B_\varphi(r)=\frac{M^2\Phi_0}{2\pi\sqrt{a^2-a_c^2}}\sum_{\sigma=\pm}\sigma M_\sigma K_1(M_\sigma r),
\end{equation}
\begin{equation}
	\label{Eq:Bz}
	B_z(r)=\frac{M^2\Phi_0}{2\pi\sqrt{a_c^2-a^2}}\sum_{\sigma=\pm}M_\sigma K_0(M_\sigma r),
\end{equation}
where $K_\alpha(x)$ are modified Bessel functions of the second kind, and  
\begin{equation}
	\label{Eq:Mpmb0t}
	2 M_{\pm}=\sqrt{a_c^2-a^2}\pm ia
\end{equation}
where $a_c=2M$. Note that despite the complex values of $M_{\pm}$, the components of the magnetic induction in Eqs.~\eqref{Eq:Bphi} and \eqref{Eq:Bz} are real. These solutions are only valid for $a<a_c$. When $a\geq a_c$,  $M_\pm$ become purely imaginary resulting in poles in Eq.~(\ref{eq:nBi}) and the integrals leading to $\vec{B}(r)$ are no longer defined~\footnote{Recall that unlike the Bessel functions $J_\alpha(z)$, $K_\alpha(z)$ is only defined for ${\rm Re} ~z>0$.}. Consequently, there is no vortex solution for  $a\geq a_c$ and one has to consider Eq. \eqref{eq:nabla2B} with a trivial vorticity ($\vec{\Omega}=0$). Therefore, the only solution for the infinite system in this case is $\vec{B}=0$, a perfect diamagnetic response. Interestingly, expressions similar to Eqs.~(\ref{eq:nabla2B}-\ref{Eq:Bz}) have arisen in the context of noncentrosymmetric  superconductors~\cite{Garaud2020,Samoilenka2020,Garaud2023}.

Due to the axion, a $\varphi$-component of the magnetic field is generated and, as a consequence, a component of the current parallel to the vortex is induced. The total current screening the vortex is thus encircling it in a helical manner, with a handedness/chirality determined by the sign of $a$. 
Figure \ref{Fig:fig-vortex-infinite} displays the magnetic induction components corresponding to the vortex solution of Eqs. (\ref{Eq:Bphi}) and (\ref{Eq:Bz}) for different values of $a$. Note that the fields start to develop more spatial structure with increasing $a$, so the Meissner screening is spatially oscillating. For $a$ closer to $a_c$ the oscillations are progressively enhanced, see Fig. \ref{Fig:fig-vortex-infinite}(c-d).

Since at $a\geq a_c$ the vortex solution breaks down, the system transitions into a superconducting phase without vortices. As at the SC phase transition the penetration depth $\lambda$ diverges and subsequently $M$ tends to zero, for any finite (and possibly small) intrinsic $a$, the regime $a\geq a_c$ is always realized close to the SC phase transition. Moreover, if the case $a\geq a_c$ is reached at any temperature, the Weyl SC only features a regime where no vortices are present --- a behavior characteristic of a type I SC.  Such a perfect diamagnetic character is also observed for a vortex solution in a finite slab of thickness $L$, with the vortex line perpendicular to the surface, see Appendix \ref{appendix:vortex-slab-sol}. There too the vortex solution does not exist in the case of $a\geq a_c$ leading to a type I London response.

\section{Meissner state}

\begin{figure}
	\begin{center}
		\subfloat[$a=0.5a_c$]{\includegraphics[width=.5\columnwidth]{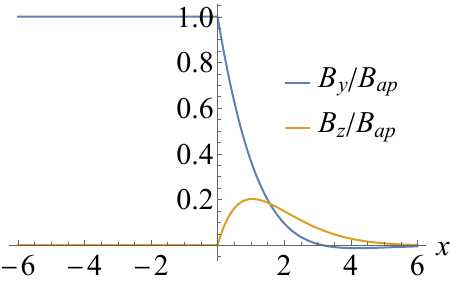}}
		\subfloat[$a=2.5a_c$]{\includegraphics[width=.5\columnwidth]{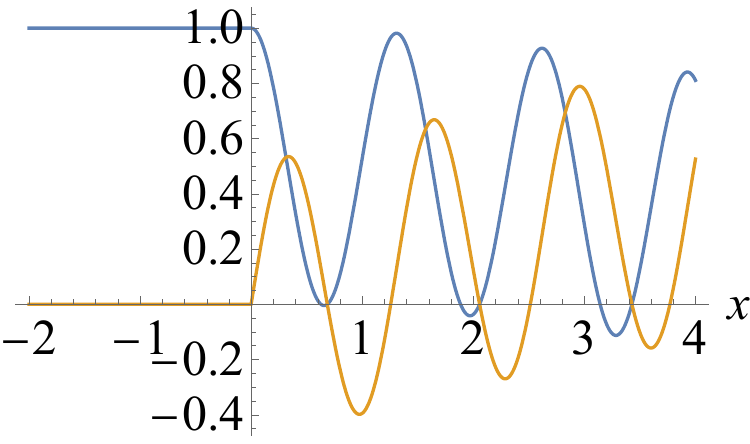}}\\
		\subfloat[$a=0.4a_c$]{\includegraphics[width=0.5\linewidth]{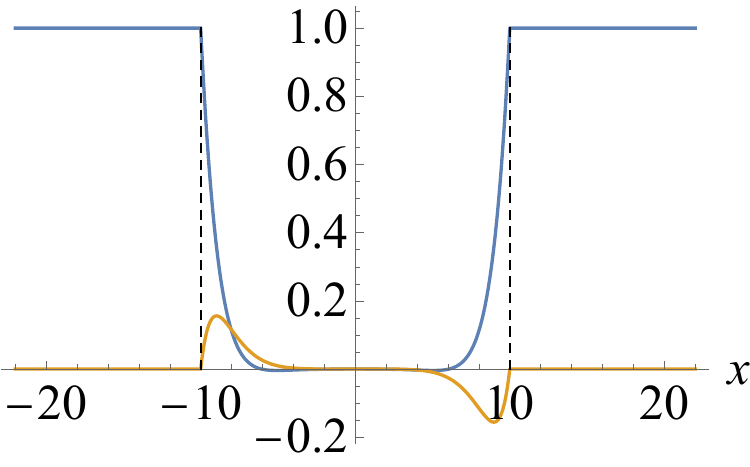}}
		\hfill
		\subfloat[$a=0.75a_c$]{\includegraphics[width=0.5\linewidth]{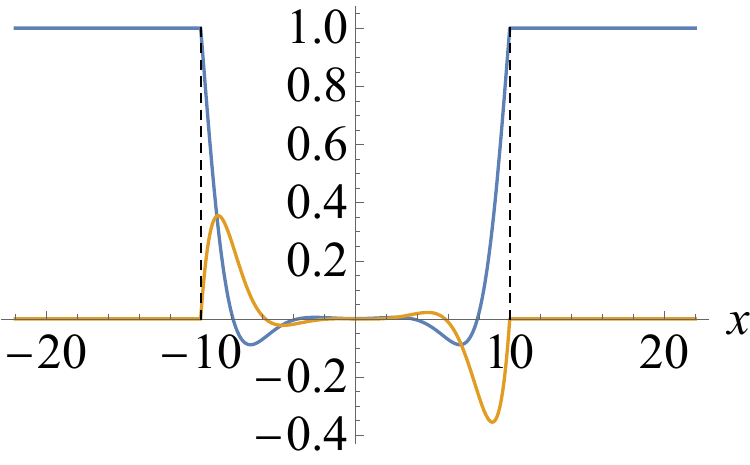}}\\
		\subfloat[$a=0.95a_c$]{\includegraphics[width=0.5\linewidth]{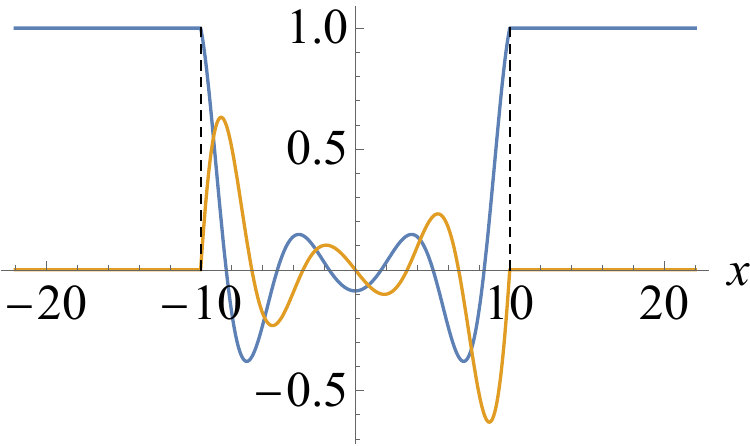}} 
		\hfill	
		\subfloat[$a=1.1a_c$]{\includegraphics[width=0.5\linewidth]{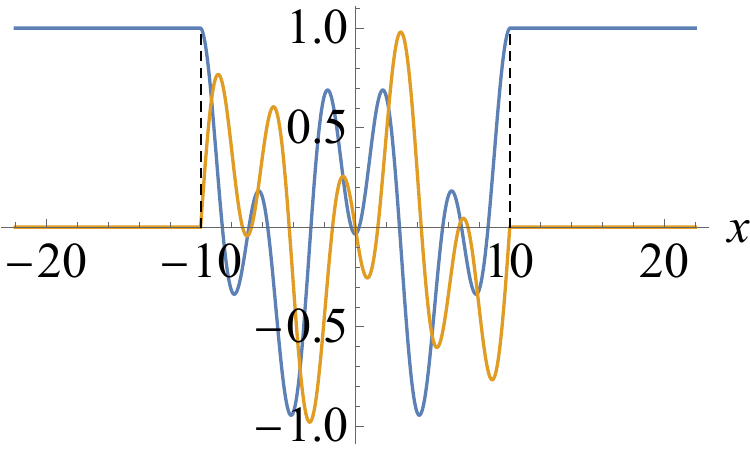}}
		\caption{Panels (a) and (b): Magnetic field components of a semi-infinite superconductor located at $x>0$ and in the presence of an applied magnetic field, $\vec{B}_{\rm ap}=B_{\rm ap}\hat{\vec{y}}$. The chiral magnetic screening occurs for $a<a_c$ (a) and is absent for $a\geq a_c$ (b). 		Panels (c)-(f): Magnetic field profiles in a finite TRI Weyl SC.  At panels (c-e) we have $a<a_c$, corresponding to $a=0.4a_c$, $a=0.75a_c$, and $a=0.95a_c$, respectively. We can see once more the onset of spatial oscillations in the magnetic induction as $a$ increases. In panel (f) $a=1.1a_c$ leading to the case where the axion coupling completely dominates over Meissner screening.}
		\label{Fig:semi-inf1} 
	\end{center}
\end{figure}

Having discussed vortex solutions, we consider now a superconducting state without vortices. Here we find a crucial difference relative to the usual London electrodynamics: the Meissner screening works differently, since application of an external magnetic field generates an additional component of the magnetic induction due to the axion term. As a first example, let us consider a {\it semi-infinite} superconductor located in the region $x>0$ in the presence of an applied magnetic field $\vec{B}_{\rm ap}=B_{\rm ap}\hat{\vec{y}}$ parallel to the surface. For this simple geometry one obtains  from Eq. \eqref{eq:nabla2B},
\begin{eqnarray}
	\label{eq:type1eq1}
	&& -\partial_x^2 B_y+M^2 B_y-a \partial_x B_z=0, \\
	&& -\partial_x^2B_z+M^2 B_z+a \partial_x B_y=0,
		\label{eq:type1eq2}
\end{eqnarray}
with the boundary conditions, $B_y(x =0)=B_{\rm ap}$, $B_y(x \rightarrow\infty)=0$,  $B_z(x =0)=0$, $B_z(x \rightarrow\infty)=0$. As with the vortex solution, there are two distinct regimes to consider: $a<a_c$ and $a\geq a_c$. The former yields the solution
	$\vec{B}(x)=B_{\rm a p} e^{-(x/2)\sqrt{a_c^2-a^2}}\hat{\vec{u}}(x)$
in terms of the unit vector, 
\begin{equation}
	\label{Eq:u}
	\hat{\vec{u}}(x)=\cos \left(ax/2\right)\hat{\vec{y}}+\sin \left(ax/2\right)\hat{\vec{z}},
\end{equation}
and one observes that the field inside the SC rotates with respect to the applied one; a chiral oscillatory feature we found in the vortex as well, for instance, illustrated in Fig.~\ref{Fig:fig-vortex-infinite}.  Thus, applying a magnetic field in the $y$-direction does not only lead to a Meissner state with an exponentially decaying $y$-component of the field, but also generates a similarly decaying field along the $z$-direction as a consequence of the axion coupling. The corresponding field profile is depicted in panel (a) of Fig.~\ref{Fig:semi-inf1} for an exemplary value of $a$ below $a_c$. Note that the axion leads to a larger penetration depth, $\widetilde{\lambda}=1/\sqrt{M^2-a^2/4}$, compared to the one in the non-Weyl SC, $\lambda=1/M$. 

For $a\geq a_c$ there is no solution that fulfills the boundary conditions $B_y(x \rightarrow\infty)=0$ and  $B_z(x \rightarrow\infty)=0$. Hence, instead of demanding the magnetic field to vanish inside the sample, we only enforce the boundary conditions at the surface $B_y(x =0)=B_{\rm ap}$,  $B_z(x =0)=0$, and require the solutions to be real, which yields
\begin{equation}
	\label{Eq:sol-semi-inf}
	\vec{B}(x)=B_{\rm a p} \cos(\sqrt{a^2-a_c^2} ~x/2)\hat{\vec{u}}(x).
\end{equation}
Remarkably, the magnetic field inside of the Weyl SC exhibits a purely oscillatory behavior illustrated in Fig.~\ref{Fig:semi-inf1}-(b) for $a=2.5a_c$. There is no magnetic field screening inside the bulk of the SC for $a\geq a_c$. Instead, it rotates around the surface normal while penetrating all the way into the bulk. 

While the solution above corresponds to the instance where the  external magnetic field is applied parallel to the surface, the Weyl SC response to the field perpendicular to the surface is rather simple.
 For $\vec{B}_{\rm ap}=B_{\rm ap}\hat{\vec{x}}$ continuity of the normal component implies that the CME current vanishes, in which case the usual London equation follows. Therefore, the magnetic induction inside the semi-infinite slab has the conventional London form, $\vec{B}=B_{\rm ap} e^{-M x}\hat{\vec{x}}$. 
This angle-dependence of the chiral magnetic field screening is quantifiable through the magnetic helicity,  $	\mathscr{H}=\int dx \vec{A}\cdot \vec{B}$, which is a measure for the twistedness of the $\vec{B}$ field. For an arbitrary directed external field, $\vec{B}_{\rm ap}=B_{\rm ap}\left(\cos\theta \vec{\hat{e}}{_\perp} + \sin\theta\vec{\hat{e}}{_\parallel}\right)$ (where  $\vec{\hat{e}}{_\perp}$ and  $\vec{\hat{e}}{_\parallel}$ are unit vectors perpendicular and parallel to the surface, respectively) the helicity per area has a sinusoidal dependence on $\theta$, where $\theta$ is the angle between the applied field and the surface normal of a Weyl SC. Hence, the conventional exponential decay for $\vec{B}_{\rm a p}$ perpendicular to the surface corresponds to vanishing helicity, while for the semi-infinite slab in a parallel field and $a<a_c$ the helicity per unit area is finite,
$
\mathscr{H}/B_{\rm ap}^2 =- \frac{2 a}{a_c^2\sqrt{a_c^2 -a^2}},
$
and diverges at $a_c$. This brings to the fore once more the important role of the critical axion coupling $a_c$ in modifying the nature of the Meissner state.

To explore further regimes bellow and above $a_c$, we consider a {\it finite slab} geometry, where the TRI Weyl SC is between two surfaces such that $|x|\leq L/2={\bar L}$. This geometry additionally allows    to compute the magnetic susceptibility,
as this requires us to determine the average magnetic induction over the system, which cannot be done easily in a convergent manner in a semi-infinite system. Having applied the boundary conditions that require $\vec{B}(x=\pm \bar{L})=\vec{B}_{\rm ap}$, we obtain, 
\begin{eqnarray}
	\label{Eq:B-slab}
	&&\hspace{-0.5cm}\vec{B}(x)=\frac{B_{\rm ap}}{\sin(\bar L \sqrt{a^2-a_c^2})}
	\nonumber\\
	&&\hspace{-0.5cm}\times\sum_{\sigma=\pm}\sigma\sin\left[\frac{\sqrt{a^2-a_c^2}}{2}\left(x+\sigma\bar L\right)\right]\hat{\vec{u}}(x-\sigma \bar L).
\end{eqnarray}
The expression above holds for any value of $a$. Figure \ref{Fig:semi-inf1}-(c-f) shows the magnetic induction profiles corresponding to Eq. (\ref{Eq:B-slab}) for increasing values of $a$ up to slightly above $a_c$. Similarly to the findings above, the slab solution features two distinct regimes for $a<a_c$ and $a\geq a_c$. The magnetic induction inside the SC rotates for any $a$, but for $a<a_c$ it is screened, while for $a\geq a_c$ the field penetrates into the whole slab.
Remarkably, in the latter regime the oscillation amplitude can get larger than the applied field. This exotic behavior in a system with a nonzero superconducting order parameter may raise a question whether the energy of such a state is bounded below. To check that the solutions for $a\geq a_c$ are physically valid, one must go beyond the linearized London theory and consider the Ginzburg-Landau energy in the London approximation (see Appendix \ref{appendix:energy}).

We determine the diamagnetic susceptibility $\chi$ via the spatial average of the magnetic induction. It is clear that axion-induced field component $B_z$ averages to zero for any $a$. Hence, the susceptibility  
\begin{equation}
	\label{Eq:sucep}
	\chi=\frac{\sqrt{a^2-a_c^2}\left[\cos({\bar L}\sqrt{a^2-a_c^2})
		-\cos(a {\bar L})\right]}{LM^2\sin({\bar L}\sqrt{a^2-a_c^2})}-\frac{1}{4\pi}
\end{equation}
is determined by the field component parallel to the applied field.
Considering the case of a large slab thickness compared to the London penetration depth, $ML\gg 1$, one finds that $\chi=-1/(4\pi)$ for all $a\leq a_c$, as can be seen from the form of $\chi=\frac{2}{M^2L^2}[1-\cos(ML)]-1/(4\pi)$ for $a\to a_c$ specifically. This implies a perfect diamagnetic response of the bulk superconductor.
  
For $a>a_c$, on the other hand, the situation is more delicate and interesting. From the numerator of the first term in Eq.~(\ref{Eq:sucep}), one observes that $\chi=-1/(4\pi)$ is reached for values of $a$ given by
\begin{equation}
	\label{Eq:amin}
	a_{\rm min}(n)=\frac{\pi n}{\bar{L}}+\frac{a_c^2\bar{L}}{4\pi n}~~~{\rm with}~~~~n\in\mathbb{N}.
\end{equation}
Interestingly, for any $a_{\rm min}(n)$ the magnetic induction inside the slab satisfies a force-free equation characteristic of a TRI Weyl semimetal (see Appendix \ref{appendix:force-free}). Thus, in this case the Weyl superconductor with $M \neq 0$ has exactly the same magnetic induction as a normal $M=0$ Weyl semimetal.
On the other hand, the susceptibility diverges for
\begin{equation}
	\label{eq:aquant}
	a_{\rm max}^2(m)=a_c^2+\left(\frac{\pi m}{\bar{L}}\right)^2~~~{\rm with}~~~~m\in\mathbb{N}.
\end{equation}
Consequently, at quantized values of the axion coupling the system becomes unstable. Exactly at this progression of critical couplings, the magnetic helicity of the slab diverges as well; see Appendix \ref{appendix:Helicity}. 
The two sets of relations characterize two distinct behaviors of the electromagnetic response to the external magnetic field. 
When $a$ takes the values $a_{\rm min}(n)$, the susceptibility $\chi\to-1/(4\pi)$ irrespective of the slab thickness $L$ in contrast to the conventional Meissner effect, where $\chi\to-1/(4\pi)$ only for sufficiently thick samples.
The values $a_{\rm max}(m)$ separate transitions between these different onsets of chiral Meissner regimes, and occur due to poles present in the Green function of the operator $(-\nabla^2+M^2+a\nablab\times)$ in the slab geometry. Hence the regime for $a>a_c$ exhibits a helimagnetic behavior, since the Green function in this case features poles at nonzero momenta.

\section{Conclusions and outlook}

As the axion term affects the properties of time-reversal invariant Weyl superconductors in a quite nontrivial manner, several distinct experimentally testable predictions follow from our results. 
In the vortex state the currents parallel to magnetic vortices are induced by the axion coupling.  In future work it will be interesting to establish how this affects the vortex lattice and its stability. Additionally, the rotating $B$-field  causes vortices close to the surface to cant with respect to the applied field, which results in magnetic stray fields outside the Weyl SC~\cite{tbp}. This is a rather intricate consequence of the emergent field components transverse to the flux line, as simple addition of a mirror vortex cannot fulfill the boundary conditions on the surface. 

We find that the the vortex becomes unstable at the critical axion coupling implying a transition from a type II to a type I superconducting state. 
Such a transition between type I and II SC in the same material is known as type 1.5 superconductivity for multiband systems~\cite{Babaev2005,Moshchalkov2009}.
As close to the SC phase transition the London penetration depth diverges, the critical axion coupling vanishes there. 
Thus for any given axion coupling intrinsic to the TRI Weyl SC material, close enough to the SC transition the system automatically enters the strong coupling regime where the vortex state becomes unstable and a type 1.5 regime may ensue.
Interestingly, studies of chiral effects in an astrophysical setting have established that for vanishing critical axion coupling a chiral plasma instability (CPI) emerges, which is strongly affected by field fluctuations~\cite{Kamada2023}.  For Weyl SCs close to $T_{\rm c}$ the CPI must then be considered on the same footing as superconducting order parameter fluctuations. The fate of this fluctuating chiral SC is an open question for future work.

In type I SCs the axion induced magnetic field component perpendicular to the applied field and parallel to the surface may be explored by surface sensitive probes, e.g. the magneto-optic Kerr effect. For strong coupling the axion renormalization of the London penetration depth may be probed experimentally. Also the periodically divergent susceptibility associated with the chiral Meissner state is a marked experimental signature. 

{\it Note added.} Recently, we became aware of the work of Stålhammar et al. \cite{stalhammar2023}, who report semi-infinite slab and cylinder solutions for $a<a_c$ similar to our results.

\begin{acknowledgments}
	We thank Volodymyr Kravchuk for stimulating discussions. We acknowledge financial support by the Deutsche Forschungsgemeinschaft (DFG, German Research Foundation), through SFB 1143 project A5 and the W{\"u}rzburg-Dresden Cluster of Excellence on Complexity and Topology in Quantum Matter-ct.qmat (EXC 2147, Project Id No. 390858490). 
\end{acknowledgments}

\appendix

\section{Vortex solution in a slab geometry}
\label{appendix:vortex-slab-sol}

The vortex solution for a slab defined in the region $|z|<L/2$ is more easily obtained by considering the differential equations for the vector potential. The London equation for the vector potential reads, 
\begin{equation}
	-\nabla^2 \vec{A}+M^2 \vec{A}=\frac{M^2 \Phi_0}{2 \pi} \frac{\hat{\varphib}}{r}-a \nablab\times\vec{A}.
\end{equation}
\begin{widetext}
	Based on the infinite vortex solution, we consider the Ansatz, 
	\begin{eqnarray}
		A_r(r,z)&=&a \frac{M^2 \Phi_0}{2 \pi} \int d p \frac{J_1(p r)}{ (p^2+M^2)^2-a^2 p^2} \frac{\partial \beta(p,z)}{\partial z},\\
		A_{\varphi}(r,z)&=&\frac{M^2 \Phi_0}{2 \pi} \int d p \frac{J_1(p r) (p^2+M^2)}{ (p^2+M^2)^2-a^2 p^2} \beta(p, z), \\
		A_z(r,z)&=&-a \frac{M^2 \Phi_0}{2 \pi} \int d p \frac{p J_0(p r)}{ (p^2+M^2)^2-a^2 p^2} \beta(p, z),	
	\end{eqnarray}
	\begin{eqnarray}
		B_r&=&-\frac{M^2 \Phi_0}{2 \pi} \int d p \frac{\left(p^2+M^2\right) J_1(p r)}{\left(p^2+M^2\right)^2-a^2 p^2} \frac{\partial \beta(p,z)}{\partial z}, \\
		B_{\varphi}&=&-\frac{a M^2 \Phi_0}{2 \pi} \int d p \frac{J_1(p r)\left[p^2 \beta-\frac{\partial^2 \beta(p,z)}{\partial z^2}\right]}{\left(p^2+m^2\right)^2-a^2 p^2}, \\
		B_z&=&\frac{M^2 \Phi_0}{2 \pi} \int d p \frac{p\left(p^2+M^2\right) J_0(p r)}{\left(p^2+M^2\right)^2-a^2 p^2} \beta(p,z),
	\end{eqnarray}
	where the function $\beta(p,z)$ is determined by application of the boundary conditions, which imposes the continuity of $\vec{A}$ and its derivatives with respect to $z$ at the interfaces $z=\pm L/2$. With this we obtain, 
	\begin{equation}
		\begin{aligned}
			\beta(p, z)
			& = \left\{\begin{array}{l}
				e^{p\left(\frac{L}{2}-z\right)} \frac{\left[\tau_1 \sinh \left(\frac{L \tau_1}{2}\right) \cos \left(\frac{L \tau_2}{2}\right)-\tau_2 \cosh \left(\frac{L \tau_1}{2}\right) \sin \left(\frac{L \tau_2}{2}\right)\right]}{p \cosh \left(\frac{L \tau_1}{2}\right) \cos \left(\frac{L \tau_2}{2}\right)+\tau_1 \sinh \left(\frac{L \tau_1}{2}\right) \cos \left(\frac{L \tau_2}{2}\right)-\tau_2 \cosh \left(\frac{L \tau_1}{2}\right) \sin \left(\frac{L \tau_2}{2}\right)}  \\
				1-\frac{p \cosh \left(\tau_{1}z\right) \cos \left(\tau_2 z\right)}{p \cosh \left(\frac{L \tau_1}{2}\right) \cos \left(\frac{L \tau_2}{2}\right)+\tau_1 \sinh \left(\frac{L\tau_1}{2}\right) \cos \left(\frac{L \tau_2}{2}\right)-\tau_2 \cosh \left(\frac{L \tau_1}{2}\right) \sin \left(\frac{L \tau_2}{2}\right)} \\
				e^{p\left(\frac{L}{2}+z\right)} \frac{\left[\tau_1 \sinh \left(\frac{L \tau_1}{2}\right) \cos \left(\frac{L \tau_2}{2}\right)-\tau_2 \cosh \left(\frac{L \tau_1}{2}\right) \sin \left(\frac{L \tau_2}{2}\right)\right]}{p \cosh \left(\frac{L \tau_1}{2}\right) \cos \left(\frac{L \tau_2}{2}\right)+\tau_1 \sinh \left(\frac{L \tau_1}{2}\right) \cos \left(\frac{L \tau_2}{2}\right)-\tau_2 \cosh \left(\frac{L \tau_1}{2}\right) \sin \left(\frac{L \tau_2}{2}\right)},
			\end{array}\right.
		\end{aligned}
	\end{equation}
	where we defined,
	\begin{equation}
		\tau_1^2(p)=\frac{1}{2}\left(\sqrt{\left(p^2+M^2\right)^2-a^2 p^2}+p^2+M^2-\frac{a^2}{2}\right),
	\end{equation}
	\begin{equation}
		\tau_2^2(p)=\frac{1}{2}\left(\sqrt{\left(p^2+M^2\right)^2-a^2 p^2}-\left(p^2+M^2-\frac{a^2}{2}\right)\right).
	\end{equation}
	We note that,
	\begin{equation}
		\tau_1(0)=\sqrt{4M^2-a^2}/2,~~~~~~~~\tau_2(0)=ia/2, 
	\end{equation}
	and thus we obtain the complex mass scales $M_\pm=\tau_1(0)\pm\tau_2(0)$ that appear in the infinite vortex solution in the main text. 
	
	Up to linear order in $a$,
	\begin{eqnarray}
		B_{\varphi}&=&-\frac{a M^2 \Phi_0}{2 \pi} \int d p \frac{J_1(p r)\left[p^2 \beta_0(p,z)-\frac{\partial^2 \beta_0(p,z)}{\partial z^2}\right]}{\left(p^2+M^2\right)^2} \\
		B_r&=&-\frac{M^2 \Phi_0}{2 \pi} \int d p \frac{ J_1(p r)}{\left(p^2+M^2\right)} \frac{\partial \beta_0(p,z)}{\partial z} \\
		B_z&=&\frac{M^2 \Phi_0}{2 \pi} \int d p \frac{p J_0(p r)}{\left(p^2+M^2\right)} \beta_0(p,z),
	\end{eqnarray}
	where the notation $\beta_0(p,z)$ refers to the function $\beta(p,z)$ with $a=0$,
	\begin{equation}
		\beta_0(p, z)=\left\{\begin{array}{l}
			\frac{ \tau e^{p\left(\frac{L}{2}-z\right)}}{ \tau+p \operatorname{coth}\left(\frac{\tau L}{2}\right)}, \quad z>\frac{L}{2} \\
			1-\frac{p \cosh \left(\tau z\right)}{p \cosh \left(\frac{\tau L}{2}\right)+ \tau \sinh \left(\frac{\tau L}{2}\right)}, \quad-\frac{L}{2}<z<\frac{L}{2} \\
			\frac{ \tau e^{p\left(\frac{L}{2}+z\right)}}{\tau+p \operatorname{coth}\left(\frac{\tau L}{2}\right)}, \quad z<-\frac{L}{2},
		\end{array}\right.
	\end{equation}
	$\tau = \sqrt{p^2 + M^2}$.
	
\section{Ginzburg-Landau energy in the London approximation}
	\label{appendix:energy}	
	One may wonder whether the energy is bound from below in the state without vortices for $a\geq a_c$. The first thing to observe here is that the calculations in the strict London regime have their limitations regarding this instability. This is more easily seen by considering the energy density associated to our London equations, 
	\begin{equation}
		\label{Eq:E1}
		\mathcal{E}=\frac{1}{8\pi}\vec{B}\cdot(\vec{B}+a\vec{A})+\frac{M^2}{8\pi}\vec{A}^2.
	\end{equation}  
	This invites to rewrite the energy density in the form, 
	\begin{equation}
		\label{Eq:E2}
		\mathcal{E}=\frac{1}{8\pi}\left(\vec{B}+\frac{a}{2}\vec{A}\right)^2+\frac{1}{32\pi}\left(a_c^2-a^2\right)\vec{A}^2.
	\end{equation}
	In the form above, the energy density is not positively defined for $a>a_c$, and eventually the energy is not even bounded from below. However, this is actually a limitation of the London regime, which can be fixed by obtaining further corrections to the London theory directly from the Ginzburg-Landau (GL) energy functional,
	\begin{equation}
		\label{Eq:GL}
		\mathcal{E}=\frac{1}{8\pi}(\vec{B}+a\vec{A})\cdot\vec{B}+|(\nablab-iq\vec{A})\psi|^2+m^2|\psi|^2+\frac{u}{2}|\psi|^4.
	\end{equation}  
	The GL equations are, 
	\begin{equation}
		\label{Eq:GL-eq1}
		\nabla^2\psi+2iq\vec{A}\cdot\nablab\psi=\left(m^2+u|\psi|^2+q^2\vec{A}^2\right)\psi,
	\end{equation}
	\begin{equation}
		\label{Eq:GL-eq2}
		\frac{1}{4\pi}\nablab\times\left(\vec{B}+a\vec{A}\right)=-2q^2|\psi|^2\vec{A}-iq(\psi^*\nablab\psi-\psi\nablab\psi^*).
	\end{equation}
	At this stage it is convenient to employ the unitary gauge where $\psi=\rho e^{i\theta}/\sqrt{2}$. In the absence of vortices we gauge away the phase, $\vec{A}\to\vec{A}+\nabla\theta/q$, to obtain Eq. (\ref{Eq:GL-eq1}) rewritten in the form, 
	\begin{equation}
		\label{Eq:GL-eq3}
		\frac{1}{\kappa^2}\nabla^2\rho=\frac{q^2\rho_0^2}{2}\left(-1+\frac{\rho^2}{\rho_0^2}+\vec{A}^2\right)\rho,
	\end{equation}
	where $\kappa^2=u/q^2$ is the square of the GL parameter, and $\rho_0^2=-2m^2/u>0$ is the mean-field order parameter squared. We have also rescaled the vector potential as $\vec{A}\to\kappa\rho_0\vec{A}/\sqrt{2}$, which does not affect Eq. (\ref{Eq:GL-eq2}).  
	Strictly speaking, the London regime corresponds to $\kappa\to\infty$ causing the LHS of Eq. (\ref{Eq:GL-eq3}) to vanish. Therefore, 
	\begin{equation}
		\rho^2=\rho_0^2(1-\vec{A}^2).
	\end{equation}
	Substituting this back into the expression for energy in Eq. \eqref{Eq:GL}, we obtain 
	\begin{equation}
		\label{Eq:GL1}
		\mathcal{E}=\frac{1}{8\pi}\left(\vec{B}+\frac{a}{2}\vec{A}\right)^2+\frac{1}{32 \pi}\left(a_c^2-a^2\right)\vec{A}^2-\frac{\rho_0^2 m^2}{4}\left(\vec{A}^2\right)^2,
	\end{equation}  	
	where we have defined $M^2=4\pi q^2\rho_0^2$ (this leads us to identify $\rho_s=\rho_0^2$ in the manuscript) and $a_c=2M$. Comparing this to the expression in Eq. \eqref{Eq:E2} that yields linearized field equations, we discover a term $\propto\left(\vec{A}^2\right)^2$ that stabilizes the energy for $a>a_c$, since $m^2<0$.

\section{Magnetic induction at $a_{\rm min}(n)$}
\label{appendix:force-free}
The Weyl superconductor with $M \neq 0$ having the same magnetic induction as a normal $M=0$ Weyl semimetal for any $a_{\rm min}(n)$ can be seen directly by recasting the London equation for the TRI Weyl SC in the form $({\rm curl}-iM_+)({\rm curl}+iM_-)\vec{B}=0$ \cite{Beltrami-stuff} with $M_\pm$ as defined in Eq. (7) of the main text. It turns out that for $a=a_{\rm min}(n)$ we have $iM_-=-2\pi n/L$ and it follows immediately from the operator decomposition just employed that there is a solution satisfying $\nablab\times\vec{B}=(2\pi n/L)\vec{B}$, which yields, $\vec{B}_n(x)=B_{\rm ap}(-1)^n[\cos(2\pi nx/L)\hat{\vec{y}}+\sin(2\pi nx/L)\hat{\vec{z}}]$.

	%\appendix
	\section{Helicity in a finite slab}
	\label{appendix:Helicity}
	
	From the magnetic induction in a slab without vortices given in Eq.~\eqref{Eq:B-slab} one can derive components of the vector potential,
	\begin{equation}
		\begin{aligned}
			A_y(x)= & \frac{B_{\rm a p}}{2 M^2 \sinh \left(\bar{L} \sqrt{a_c^2-a^2}\right) } \times\\
			\times \sum_{\sigma=\pm} \sigma & \left[\sqrt{a_c^2-a^2} \sin \left(\frac{a}{2}\left(x-\sigma \bar{L} \right)\right) \cosh \left(\frac{x+\sigma\bar{L} }{2} \sqrt{a_c^2-a^2}\right)+a \cos \left(\frac{a}{2}\left(x+\sigma \bar{L} \right)\right) \sinh \left(\frac{x-\sigma \bar{L} } {2} \sqrt{a_c^2- a^2}\right)\right],
		\end{aligned}
	\end{equation}
	\begin{equation}
		\begin{aligned}
			A_z(x)= & \frac{B_{\rm a p}}{2 M^2 \sinh \left(\bar{L} \sqrt{a_c^2-a^2}\right) } \times\\
			\times \sum_{\sigma=\pm} \sigma & \left[\sqrt{a_c^2-a^2} \cos \left(\frac{a}{2}\left(x+\sigma\bar{L} \right)\right) \cosh \left(\frac{x-\sigma \bar{L} }{2} \sqrt{a_c^2-a^2}\right)+a \sin \left(\frac{a}{2}\left(x+\sigma \bar{L} \right)\right) \sinh \left(\frac{x-\sigma \bar{L} }{2} \sqrt{a_c^2 -a^2}\right)\right].
		\end{aligned}
	\end{equation}
	From the expressions for magnetic field components and the vector potential we can calculate total helicity per area by integrating over the sample's thickness,
	\begin{eqnarray}
		\mathscr{H}	& =&\int_{-\bar{L}}^{\bar{L}} dx\vec{A}\cdot \vec{B}=\frac{B_{\rm a p}^2}{2 M^2 \sqrt{a_c^2-a^2} \sinh ^2\left(\bar{L} \sqrt{a_c^2-a^2}\right)} \times \\
		& \times&\left[-2a \bar{L} \sqrt{a_c^2-a^2} \cos \left(a \bar{L}\right) \cosh \left(\bar{L} \sqrt{ a_c^2-a^2}\right)+2a \bar{L} \sqrt{a_c^2- a^2}-a\sinh \left(2\bar{L}\sqrt{ a_c^2-a^2}\right)\right.\nonumber \\
		&+&\left.  \left(2 a \cos \left(a \bar{L}\right)+2\left(a^2-a_c^2\right) \bar{L} \sin \left(a\bar{L}\right)\right) \sinh \left(\bar{L} \sqrt{a_c^2-a^2}\right)\right]. \nonumber
	\end{eqnarray}
	Similarly to magnetic susceptibility mentioned in the main text, the helicity is divergent for values of $a$ that satisfy the relation in Eq. (14) of the main text.

\end{widetext}

%===========================

\bibliography{weylSC}

\end{document}